# Collaborative Filtering Approach to Link Prediction

Yan-Li Lee, Tao Zhou

CompleX Lab, University of Electronic Science and Technology of China, Chengdu 611731,

People's Republic of China.

**Abstract**. Link prediction is a fundamental challenge in network science. Among various methods, local similarity indices are widely used for their high cost-performance. However, the performance is less robust: for some networks local indices are highly competitive to state-of-the-art algorithms while for some other networks they are very poor. Inspired by techniques developed for recommender systems, we propose an enhancement framework for local indices based on collaborative filtering (CF). Considering the delicate but important difference between personalized recommendation and link prediction, we further propose an improved framework named as self-included collaborative filtering (SCF). The SCF framework significantly improved the accuracy and robustness of well-known local indices. The combination of SCF framework and a simple local index can produce an index with competitive performance and much lower complexity compared with elaborately-designed state-of-the-art algorithms.

## I. Introduction

Link prediction is a paradigmatic problem in network science that attempts to uncover missing links or predict future links [1][2], which has already found many theoretical and practical applications, such as evaluation of evolving models [3][4], reconstruction of networks [5][6], recommendation of friends and products [7][8], inference of biological interactions [9][10], and so on.

Various methods for link prediction are proposed [2][11][12][13][14], including similarity-based algorithms [15][16], maximum likelihood methods [17][18][19], probabilistic models [20], latent space models [21][22], deep learning [23][24], matrix completion [25], and so on. In a similarity-based algorithm, each node pair $(x, y)$ is assigned a similarity score $S_{xy}$ and an unobserved link with higher score is assumed to be more likely to exist. If $S_{xy}$ only depends on local topology surrounding $x$ and $y$, we call it a local similarity index. Representative local similarity indices include common neighbors (CN) index [15], Adamic-Adar (AA) index [26], resource allocation (RA) index [16][27], local path (LP) index [16][28], local naïve Beyas (LNB) index [29], local random walk (LRW) index [30], Cannistraci-Hebb (CH) index [31], and many others [32][33].

Local similarity indices are widely applied for their simplicity, competitive performance and low complexity [15][16]. However, the performance of local similarity indices is not robust: for some networks they are highly competitive to state-of-the-art algorithms while for some other networks they are very poor [34][35]. We have noticed that in recommender systems, the interest of a target user on a product is usually estimated by some other users with similar purchase records to the target user. Such so-called collaborative filtering techniques [36] inspires us to propose a general framework that can considerably improve the accuracy and robustness of local similarity indices.

## II. Methods

This paper reports experimental results associated with three representative indices, and results for others are similar. The first one is the CN index [15] that simply counts the number of common neighbors as

$$S_{xy}^{CN} = |\Gamma_x \cap \Gamma_y|, \tag{1}$$

where $\Gamma_x$ and $\Gamma_y$ are sets of neighbors of nodes $x$ and $y$. The second one is the RA index [16] that is obtained by weakening the weights of large-degree common neighbors according to a resource-allocation dynamics [17], say

$$S_{xy}^{RA} = \sum_{z \in \Gamma_x \cap \Gamma_y} \frac{1}{k_z}, \tag{2}$$

where $k_z$ is the degree of node $z$. The last one is the Cannistraci resource allocation (CRA) index [32], which is an extension of RA index considering the local community paradigm, as

$$S_{xy}^{CRA} = \sum_{z \in \Gamma_x \cap \Gamma_y} \frac{|\gamma_z|}{k_z}, \tag{3}$$

where $\gamma_z$ is the subset of $z$'s neighbors that are also common neighbors of $x$ and $y$.

Before the description of the proposed enhancement framework, we briefly introduce the idea of collaborative filtering in recommender systems [8][36]. Generally speaking, a recommender system can be characterized by a user-product bipartite network [37], where a user $i$ is connected with a product $\alpha$ if $i$ has purchased $\alpha$. In the simplest version of user-based CF, the interest of $i$ on $\alpha$ can be written as

$$I_{i\alpha} = \sum_{j \in \Lambda_i} S_{ij} A_{j\alpha}, \tag{4}$$

where $\Lambda_i$ is the set of users whose purchase similarity to $i$ is larger than a preseted threshold, and

$A$ is the adjacency matrix with $A_{j\alpha} = 1$ if $j$ has purchased $\alpha$ and $A_{j\alpha} = 0$ otherwise. The binary adjacency matrix $A$ is usually replaced by a weighted one with the weight of a link being the number of purchases, the logarithm of purchased money, or other alternatives. Inspired by the user-based CF, a given similarity index $S$ can be enhanced in a way similar to Eq. (4) as

$$\dot{S}_{xy} = \sum_z A_{xz} S_{zy} + \sum_z A_{yz} S_{zx}, \tag{5}$$

where $A$ is the adjacency matrix of the target network. The difference between Eq. (4) and Eq. (5) lies in two aspects. Firstly, in Eq. (4), users with similarity larger than a threshold constitute the *neighborhood* of user $i$ (i.e., $\Lambda_i$) while in Eq. (5) the neighborhood of a user $x$ is determined by the observed links (i.e., $\Gamma_x$ or $\{z|A_{xz} = 1\}$). Secondly, for link prediction, we should simultaneously consider the neighborhoods of $x$ and $y$. Obviously, Eq. (5) can be rewritten in a matrix form as

$$\dot{S} = AS + (AS)^T \tag{6}$$

and $\dot{S}$ is a symmetry matrix. Given an arbitrary similarity index $S$, we can obtain an enhanced similarity index $\dot{S}$ via Eq. (6). As it is inspired by the user-based CF, we call it a CF-based enhancement framework for similarity index.

In classical recommender systems, we are not allowed to recommend a product $\alpha$ to user $i$ if $i$ has already purchased $\alpha$. It is because in some scenarios users are unlikely to repurchase or revisit products (e.g., watching movies [38]), and to recommend known products is not informative [39]. Therefore, the purchase records of the target user are excluded in Eq. (4) for recommender systems, namely $i \notin \Lambda_i$. In contrast, such constrain is unnecessary in link prediction. That is to say, the direct similarity between two nodes should be considered in estimating the existence likelihood of the corresponding link. Accordingly, we further propose an enhancement framework

called self-included collaborative filtering (SCF) as

$$\ddot{S} = (A + I)S + [(A + I)S]^T, \qquad (7)$$

where $I$ is the identity matrix.

## III.   Results

Consider a simple network $G(V, E)$, where $V$ and $E$ are sets of nodes and links, the directionalities and weights of links are ignored, and multiple links and self-connections are not allowed. We assume that there are some missing links in the set of nonobserved links and the task of link prediction is to find out those missing links. To test the algorithm's accuracy, the observed links, $E$, is randomly divided into two parts: the training set $E^T$ is treated as known information, while the probe set $E^P$ is used for algorithm evaluation and no information in $E^P$ is allowed to be used for prediction. We adopt the area under the receiver operating characteristic curve (AUC) [40] as the metric for algorithms' performance. AUC value can be interpreted as the probability that a randomly chosen missing link in $E^P$ is assigned a higher existence likelihood (i.e., similarity score) than a randomly chosen link in the set of nonobserved links $U \backslash E$, where $U$ is the universal set containing all $|V|(|V|-1)/2$ potential links. Among $n$ comparisons, if there are $n_1$ times the missing link having higher score and $n_2$ times the two having same score, the AUC value is

$$AUC = \frac{n_1 + 0.5 n_2}{n}. \qquad (8)$$

If all similarity scores are generated from an independent and identical distribution, the AUC value should be about 0.5. Therefore, the degree to which the value exceeds 0.5 indicates how better the algorithm performs than pure chance. Among many candidate metrics [41], we choose AUC because link prediction is a typical imbalance learning task and AUC is non-parametric and very suitable for imbalance learning.

To test the algorithmic performance, 45 real networks from disparate fields are considered in this paper, including (1) FWF [42]—the predator-prey network of animals in ecosystem of Coastal bay in Florida Bay in the dry season; (2) FWE [43]—the predator-prey network of animals in Everglades Graminoids in the west season; (3) FWM [44]—the predator-prey network of animals in Mangrove Estuary during the wet season; (4) SciM [45]—the citation network for papers that published in *Scientometrics*; (5) SmaG [45]—the citation network of citations to papers of Small & Griffith and Descendants; (6) Arnet [46]—the citation network for papers from 10 scientific fields, including data mining, databases systems, information retrieval, etc; (7) SOM [45]—the citation network for papers that cited the book entitled "Self-Organizing Maps"; (8) SMW [45]—the citation network for papers that cited the paper entitled "The small world problem" by Stanley Milgram; (9) ARE [47]—the email network between members at the University Rovira i Virgili; (10) RAD [47]—the email communication network between employees of a mid-sized manufacturing company; (11) DNC [47]—the email communication network in the 2016 Democratic Committee email leak; (12) OPS [47]—the message network between students of an online community from the University of California; (13) HFR [47]—the friendship network among users of the website hamsterster.com; (14) HFU [47]—the network of friendships and family relationships between users of the website hamsterster.com; (15) AH [47]—the friendship network between students obtained from a survey that took place in 1994-1995; (16) RH [47]—the friendship network between residents living in a residence hall inside the Australian National University campus; (17) HG [47]—the contact network of people measured by carried wireless devices; (18) IF2009 [47]—the face-to-face contact network between attendees of the exhibition INFECTIOUS:STAY AWAY in 2009 at the Science Gallery in Dublin, in which a link between two attendees is generated if the two have a face-to-face contact lasting at least 20 seconds; (19) H2009 [47]—the face-to-face contact network between attendees of the ACM Hypertext 2009 conference, similar to IF2009; (20) Wiki [48]—the voting network of users from the English Wikipedia in admin elections; (21) WR [49]—the religious social network in which links denote the follower-followee relationship in *weibo* (directions of links are ignored); (22) PH [47]—the innovation spreading network in which each node denotes a physician and each link denotes the friendship or one discussion between two physicians; (23) MR [50]—the rating network in which links denote ratings of users to movies; (24) NS [51]—the collaboration network between authors working on network theory and experiment; (25) BG [47]—the hyperlink network among blogs in the context of the 2004 US election; (26) GFA [52]—the gene functional association network in C.elegans; (27) CGT [53]—the genetic interaction network in C.elegans; (28) CLC [53]—a medium-scale protein-protein interaction network in C.elegans; (29) Bm13 [33]—the protein-protein interaction network in lit-bm-13; (30) FG [47]—the protein-protein interaction network in species of Homo sapiens; (31) DLC [53]—a medium-scale protein-protein interaction network in D.melanogaster; (32) SCC [53]—the protein co-citation network in S.cerevisiae; (33) Fly [53]—the network of cortical areas in brain of fly, where each link denotes a fiber tract between two cortical areas; (34) FTB [45]—the trading network of players of the national soccer teams among 35 countries; (35) WTN [54]—the world trading network of miscellaneous manufactures of metal among 80 countries in 1994; (36) UST [47]—the air transportation network of US; (37) ATC [47]—the network of airports or service centers, where each link denotes a preferred route between two nodes; (38) FT [47]—the network between airports of the world, in which each node denotes an airport and two nodes have a link if there exists at least one direct flight between them; (39) ER [47]—the international E-road network between cities in Europe, where each link denotes a road between two cities; (40) USPG [47]—the network between electrical equipments (including generators, transformators and substations) of the western States of the USA, where each link denotes a power supply line between two equipments; (41) S208 [55]—the electronic circuit network of s208, in which nodes represent logic gates or flip-flops, and links denote electronic transmission paths between nodes; (42) S420 [55]—the electronic circuit network of s420, similar to S208; (43) S838 [55]—the electronic circuit network of s838, similar to S208 (s208, s420 and s838 are three circuits in the circuit benchmarks released in ISCAS'89 [55]); (44) BB [47]—the co-occurrence network for nouns of the King James Version of the Bible; (45) JA [56]—the word-adjacency network of a Japanese text from "The Tale of Genji". The elementary statistics of those networks are reported in Table 1.

Each of the three representative local indices can be enhanced by Eq. (6) and Eq. (7), so that we can obtain 9 indices for comparison. Table 2 reports the results on the 45 real networks subject to AUC. To draw a clear picture, we calculate the within-category and global winning rates ($R_c$ and $R_g$): the former compares indices in each category and the latter compares all the 9 indices. In the calculation, we compare relevant indices on each of the 45 networks. In each comparison, the best-performed index gets score 1 and all others get 0. If two indices are equally best, they both get 0.5. The case with multiple winners is similar. The winning rate of an index is its total score divided by the number of comparisons (i.e., 45). According to the average AUC and winning rates, the CF framework can largely improve the performance of well-known local indices and then the SCF framework can still significantly improve the performance of CF framework. The RA category performs overall best, and the relatively poor performance of CRA category may be resulted from some networks without local community links.

**Table 1**: Structural statistics of the 45 real networks. $N$ and $M$ are the number of nodes and links, and $\langle k \rangle$, $\langle c \rangle$, $\langle l \rangle$ and $r$ are average degree, average clustering coefficient [45], average shortest path length and assortative coefficient [57]. For a very few unconnected networks, we only consider the largest component to calculate $\langle l \rangle$.

| Network | $N$ | $M$ | $\langle k \rangle$ | $\langle c \rangle$ | $\langle l \rangle$ | $r$ |
|---|---|---|---|---|---|---|
| FWF | 128 | 2106 | 32.91 | 0.33 | 1.77 | -0.10 |
| FWE | 69 | 880 | 25.51 | 0.55 | 1.64 | -0.27 |
| FWM | 97 | 1445 | 29.79 | 0.46 | 1.69 | -0.15 |
| SciM | 2729 | 10399 | 7.62 | 0.17 | 4.18 | -0.04 |
| SmaG | 1024 | 4916 | 9.60 | 0.31 | 2.98 | -0.19 |
| Arnet | 2550 | 5310 | 4.16 | 0.22 | 5.29 | 0.01 |
| SOM | 3772 | 12718 | 6.74 | 0.25 | 3.67 | -0.12 |
| SMW | 233 | 994 | 8.53 | 0.56 | 2.37 | -0.30 |
| ARE | 1133 | 5451 | 9.62 | 0.22 | 3.61 | 0.08 |
| RAD | 167 | 3250 | 38.92 | 0.59 | 1.97 | -0.30 |
| DNC | 2029 | 4384 | 4.32 | 0.22 | 3.37 | -0.31 |
| OPS | 1899 | 13838 | 14.57 | 0.11 | 3.06 | -0.19 |
| HFR | 1858 | 12534 | 13.49 | 0.14 | 3.45 | -0.09 |
| HFU | 2426 | 16631 | 13.71 | 0.54 | 3.59 | 0.02 |
| AH | 2539 | 10455 | 8.24 | 0.15 | 4.56 | 0.25 |
| RH | 217 | 1839 | 16.95 | 0.36 | 2.39 | 0.10 |
| HG | 274 | 2124 | 15.50 | 0.63 | 2.42 | -0.47 |
| IF2009 | 410 | 2765 | 13.49 | 0.46 | 3.63 | 0.23 |
| H2009 | 113 | 2196 | 38.87 | 0.53 | 1.66 | -0.12 |
| Wiki | 7115 | 100762 | 28.32 | 0.14 | 3.25 | -0.08 |
| WR | 6875 | 64712 | 18.83 | 0.3 | 3.49 | -0.24 |
| PH | 241 | 923 | 7.66 | 0.22 | 2.59 | -0.08 |
| MR | 1682 | 94834 | 112.76 | 0.36 | 2.16 | -0.19 |
| NS | 1461 | 2742 | 3.75 | 0.74 | 6.04 | -0.08 |
| BG | 1224 | 16715 | 27.31 | 0.32 | 2.74 | -0.22 |
| GFA | 297 | 2148 | 14.46 | 0.29 | 2.45 | -0.16 |
| CGT | 924 | 3239 | 7.01 | 0.59 | 3.73 | -0.19 |
| CLC | 1387 | 1648 | 2.38 | 0.08 | 7.92 | -0.26 |
| Bm13 | 3391 | 4388 | 2.59 | 0.07 | 6.61 | -0.02 |
| FG | 2239 | 6432 | 5.75 | 0.04 | 3.84 | -0.33 |
| DLC | 658 | 1129 | 3.43 | 0.12 | 6.61 | -0.19 |
| SCC | 2223 | 34879 | 31.38 | 0.41 | 2.63 | -0.15 |
| Fly | 1781 | 8911 | 10.01 | 0.26 | 2.91 | -0.09 |
| FTB | 35 | 118 | 6.74 | 0.27 | 2.13 | -0.26 |
| WTN | 80 | 875 | 21.88 | 0.75 | 1.72 | -0.39 |
| UST | 332 | 2126 | 12.81 | 0.62 | 2.74 | -0.21 |
| ATC | 1266 | 2408 | 3.80 | 0.07 | 5.93 | -0.02 |
| FT | 3425 | 19256 | 11.24 | 0.49 | 4.1 | -0.01 |
| ER | 1174 | 1417 | 2.41 | 0.02 | 18.4 | 0.09 |
| USPG | 4941 | 6594 | 2.67 | 0.08 | 18.99 | 0.00 |
| S208 | 122 | 189 | 3.10 | 0.06 | 4.93 | -0.002 |
| S420 | 252 | 399 | 3.17 | 0.06 | 5.81 | -0.01 |
| S838 | 512 | 819 | 3.20 | 0.05 | 6.86 | -0.03 |
| BB | 1773 | 9131 | 10.30 | 0.71 | 3.38 | -0.05 |
| JA | 2704 | 7998 | 5.92 | 0.22 | 3.08 | -0.26 |

As the network sparsity is a big challenge in both personalized recommendation and link prediction [41][58], we test the algorithms' robustness by varying the size of training set from 50% to 95%. As shown in Figure 1, the enhanced CN indices, either by Eq. (6) or Eq. (7), are remarkably more stable than the original CN index and perform spectacularly well when training sets contain fewer links. Results for the other two categories are similar.

**Table 2**: AUC of the 9 considered indices on the 45 real networks. Each result is averaged over 100 independent runs with probe set containing 10% links. The best-performed result for each network, the highest average AUC value and the highest winning rates are emphasized in bold.

| Network | $S^{CN}$ | $\dot{S}^{CN}$ | $\ddot{S}^{CN}$ | $S^{RA}$ | $\dot{S}^{RA}$ | $\ddot{S}^{RA}$ | $S^{CRA}$ | $\dot{S}^{CRA}$ | $\ddot{S}^{CRA}$ |
|---|---|---|---|---|---|---|---|---|---|
| FWF | 0.6005 | 0.8144 | 0.8100 | 0.6047 | 0.8359 | 0.8322 | 0.6375 | **0.8500** | 0.8482 |
| FWE | 0.6765 | 0.8457 | 0.8417 | 0.6922 | 0.8617 | 0.8582 | 0.7003 | **0.8889** | 0.8861 |
| FWM | 0.7009 | 0.8229 | 0.8209 | 0.7066 | 0.8398 | 0.8373 | 0.7345 | 0.8462 | **0.8466** |
| SciM | 0.7884 | 0.8833 | 0.8934 | 0.7892 | 0.8890 | **0.8996** | 0.6093 | 0.7396 | 0.7525 |
| SmaG | 0.8392 | 0.8743 | 0.8792 | 0.8484 | 0.8980 | **0.9031** | 0.7145 | 0.8094 | 0.8220 |
| Arnet | 0.7851 | 0.8320 | 0.8504 | 0.7855 | 0.8338 | **0.8527** | 0.6223 | 0.7015 | 0.7160 |
| SOM | 0.8138 | 0.8804 | 0.8905 | 0.8181 | 0.8903 | **0.9018** | 0.6435 | 0.7815 | 0.7920 |
| SMW | 0.8493 | 0.8843 | 0.8854 | 0.8748 | 0.9041 | **0.9074** | 0.8019 | 0.8201 | 0.8315 |
| ARE | 0.8437 | 0.8825 | 0.8900 | 0.8449 | 0.8922 | **0.9002** | 0.7002 | 0.8073 | 0.8211 |
| RAD | 0.9101 | 0.9098 | 0.9096 | 0.9166 | 0.9157 | 0.9151 | **0.9188** | 0.8836 | 0.8876 |
| DNC | 0.7962 | 0.7867 | 0.7867 | **0.7989** | 0.7913 | 0.7928 | 0.7445 | 0.7636 | 0.7657 |
| OPS | 0.7694 | 0.8977 | 0.8983 | 0.7747 | 0.9057 | **0.9070** | 0.6385 | 0.8793 | 0.8808 |
| HFR | 0.8045 | 0.9350 | 0.9359 | 0.8077 | 0.9525 | **0.9541** | 0.6538 | 0.9100 | 0.9102 |
| HFU | 0.9632 | 0.9490 | 0.9558 | 0.9673 | 0.9641 | **0.9721** | 0.9192 | 0.9165 | 0.9424 |
| AH | 0.7678 | 0.8186 | 0.8428 | 0.7682 | 0.8205 | **0.8441** | 0.5876 | 0.6282 | 0.6512 |
| RH | 0.8863 | 0.8384 | 0.8464 | **0.8933** | 0.8606 | 0.8680 | 0.8464 | 0.8012 | 0.8335 |
| HG | 0.9315 | 0.9324 | 0.9319 | 0.9322 | **0.9370** | 0.9358 | 0.9204 | 0.9007 | 0.9007 |
| IF2009 | 0.9378 | 0.9382 | 0.9422 | 0.9422 | 0.9491 | **0.9544** | 0.8650 | 0.9071 | 0.9212 |
| H2009 | 0.7748 | 0.7673 | 0.7670 | 0.7803 | 0.7692 | 0.7690 | **0.7825** | 0.7447 | 0.7500 |
| Wiki | 0.9268 | 0.9535 | 0.9537 | 0.9275 | 0.9588 | **0.9592** | 0.8814 | 0.9528 | 0.9531 |
| WR | 0.9246 | 0.9630 | 0.9632 | 0.9298 | 0.9681 | **0.9689** | 0.8712 | 0.9536 | 0.9547 |
| PH | 0.8394 | 0.8972 | 0.9071 | 0.8414 | 0.9016 | **0.9115** | 0.6511 | 0.7428 | 0.7763 |
| MR | 0.9036 | 0.9202 | 0.9208 | 0.9029 | **0.9250** | 0.9248 | 0.9059 | 0.9211 | 0.9211 |
| NS | 0.9363 | 0.8736 | 0.9401 | 0.9363 | 0.8742 | **0.9410** | 0.8109 | 0.6887 | 0.8198 |
| BG | 0.9186 | 0.9312 | 0.9315 | 0.9230 | 0.9398 | **0.9408** | 0.8943 | 0.9223 | 0.9234 |
| GFA | 0.8518 | 0.8461 | 0.8517 | 0.8706 | 0.8758 | **0.8810** | 0.7718 | 0.8179 | 0.8355 |
| CGT | 0.9244 | 0.9029 | 0.9213 | 0.9348 | 0.9279 | **0.9499** | 0.7508 | 0.8004 | 0.8231 |
| CLC | 0.5847 | 0.6292 | 0.6474 | 0.5860 | 0.6295 | **0.6485** | 0.5274 | 0.5266 | 0.5373 |
| Bm13 | 0.5914 | 0.6636 | 0.6832 | 0.5913 | 0.6643 | **0.6833** | 0.5142 | 0.5286 | 0.5370 |
| FG | 0.5505 | 0.8559 | 0.8526 | 0.5532 | **0.8584** | 0.8579 | 0.5094 | 0.7007 | 0.7012 |
| DLC | 0.6277 | 0.8030 | 0.8222 | 0.6285 | 0.8051 | **0.8249** | 0.5140 | 0.6201 | 0.6312 |
| SCC | 0.9475 | 0.9288 | 0.9310 | **0.9596** | 0.9557 | 0.9578 | 0.9281 | 0.9151 | 0.9248 |
| Fly | 0.8647 | 0.8788 | 0.8802 | 0.8729 | 0.8919 | **0.8941** | 0.7832 | 0.8619 | 0.8659 |
| FTB | 0.6506 | 0.7670 | 0.7580 | 0.6461 | **0.7798** | 0.7701 | 0.6739 | 0.7449 | 0.7604 |
| WTN | 0.8591 | 0.8783 | 0.8798 | 0.8954 | 0.8954 | **0.8980** | 0.8912 | 0.8193 | 0.8438 |
| UST | 0.9349 | 0.8951 | 0.9013 | **0.9517** | 0.9206 | 0.9298 | 0.9211 | 0.8683 | 0.8826 |
| ATC | 0.6099 | 0.6974 | 0.7170 | 0.6099 | 0.6985 | **0.7186** | 0.5102 | 0.5252 | 0.5304 |
| FT | 0.9479 | 0.9332 | 0.9422 | 0.9510 | 0.9458 | **0.9561** | 0.8932 | 0.8991 | 0.9047 |
| ER | 0.5249 | 0.5360 | 0.5563 | 0.5247 | 0.5359 | **0.5565** | 0.5000 | 0.5000 | 0.5000 |
| USPG | 0.5879 | 0.5966 | 0.6395 | 0.5886 | 0.5963 | **0.6399** | 0.5141 | 0.5083 | 0.5175 |
| S208 | 0.5307 | 0.5261 | 0.5431 | 0.5314 | 0.5286 | **0.5459** | 0.4999 | 0.4998 | 0.4997 |
| S420 | 0.5397 | 0.5549 | 0.5775 | 0.5401 | 0.5563 | **0.5795** | 0.5000 | 0.4999 | 0.4999 |
| S838 | 0.5489 | 0.5773 | 0.6071 | 0.5490 | 0.5779 | **0.6086** | 0.5000 | 0.4999 | 0.4999 |
| BB | 0.9720 | 0.8999 | 0.9223 | **0.9812** | 0.9287 | 0.9592 | 0.9002 | 0.8283 | 0.8916 |
| JA | 0.7586 | 0.8518 | 0.8500 | 0.7619 | **0.8596** | 0.8591 | 0.6497 | 0.8051 | 0.8054 |
| $R_c$ | 24.4% | 15.6% | **60%** | 15.5% | 17.8% | **66.7%** | 25.5% | 7.8% | **66.7%** |
| $R_g$ | 0.0% | 0.0% | 0.0% | 11.1% | 11.1% | **66.7%** | 4.45% | 4.45% | 2.2% |
| ⟨AUC⟩ | 0.7844 | 0.8279 | **0.8373** | 0.7897 | 0.8380 | **0.8482** | 0.7180 | 0.7673 | **0.7800** |

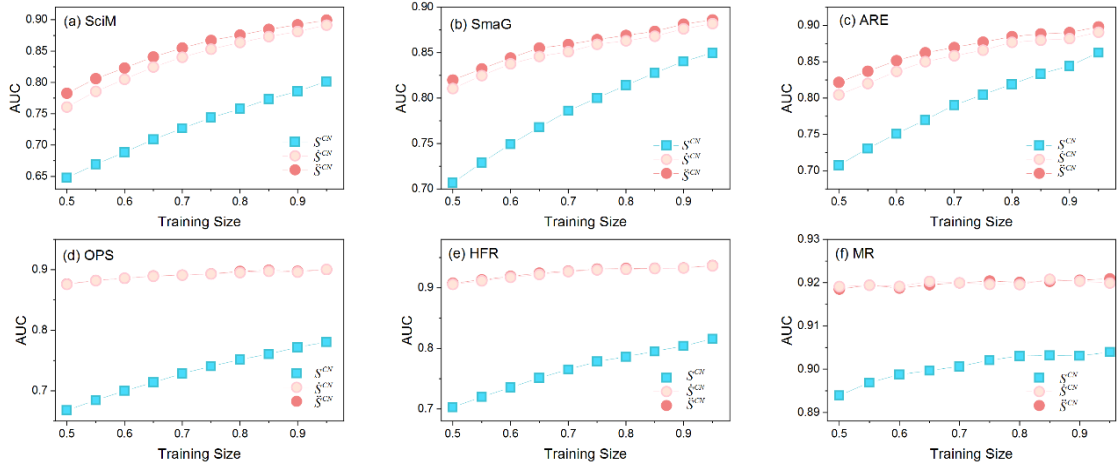

**Figure 1**: AUC of the three indices in the CN category for different training size on six selected real networks. Results are averaged over 100 independent runs. Results for the other two categories and other networks are similar.

As the SCF-enhanced indices perform overall best compared with representative local indices, we further compare them with three global algorithms, including the structural perturbation method (SPM) [59], the linear optimization (LO) [60] and the Katz index [61]. SPM randomly selects some links $\Delta E$ to construct the perturbed matrix $\Delta A$, and the background matrix $A^R = A - \Delta A$. The predicted matrix can be obtained by perturbing $A^R$ by $\Delta A$, as [59]

$$\tilde{A} = \sum_{d=1}^{N} (\lambda_d + \Delta\lambda_d) x_d x_d^T, \qquad (9)$$

where $\lambda_d$ and $x_d$ are the $d$th eigenvalue and the corresponding orthogonal and normalized eigenvector of $A^R$, and $\Delta\lambda_d = \frac{x_d^T \Delta A x_d}{x_d^T x_d}$. The entries of $\tilde{A}$ are considered as existence likelihoods of links. The final predicted matrix $\langle \tilde{A} \rangle$ is obtained by averaging over 30 independent selections of $\Delta E$. LO assumes that the existence likelihood of a link is a linear summation of contributions of all its neighbors. By solving a corresponding optimization function, the analytical expression of the similarity matrix is [60]

$$S^{LO} = \alpha A (\alpha A^T A + I)^{-1} A^T A, \qquad (10)$$

where $\alpha$ is a free parameter. The Katz index considers all possible paths connecting nodes $x$ and $y$

with exponentially damping weight, as

$$S_{xy}^{Katz} = \beta A_{xy} + \beta^2 (A^2)_{xy} + \beta^3 (A^3)_{xy} + \cdots = (I - \beta A)^{-1} - I, \tag{11}$$

where $\beta$ is a free parameter.

The results of the three SCF-enhanced indices and the three global algorithms are reported in Table 3. To our surprise, $\ddot{S}^{RA}$ still performs overall best, and LO is the second runner. All the three global algorithms refer to the matrix inversion operator and have time complexity $O(N^3)$, while the time complexities of $\ddot{S}^{CN}$, $\ddot{S}^{RA}$ and $\ddot{S}^{CRA}$ are $O(N\langle k\rangle^3)$, $O(N\langle k\rangle^3)$ and $O(N\langle k\rangle^4)$ (given an original index of time complexity $O(Q)$, its CF-enhanced and SCF-enhanced indices are of the same time complexity $O(Q\langle k\rangle)$). In large-scale and sparse networks, $\langle k\rangle \ll N$, and thus the CF-enhanced and SCF-enhanced indices are computationally more efficient than common global indices.

**Table 3**: AUC of the 6 considered algorithms on 6 simple networks. Each result is averaged over 100 independent runs with probe set containing 10% random links. The best-performed result for each network and the highest average AUC value are emphasized in bold.

| Network | $\ddot{S}^{CN}$ | $\ddot{S}^{RA}$ | $\ddot{S}^{CRA}$ | SPM | LO | $S^{Katz}$ |
|---|---|---|---|---|---|---|
| SciMet | 0.8934 | 0.8996 | 0.7525 | 0.8681 | 0.8731 | **0.9062** |
| SmaGri | 0.8792 | **0.9031** | 0.8220 | 0.8603 | 0.8842 | 0.8853 |
| ARE | 0.8900 | **0.9002** | 0.8211 | 0.8702 | 0.8901 | 0.8921 |
| OPS | 0.8983 | 0.9070 | 0.8808 | 0.8773 | **0.9083** | 0.8753 |
| HFR | 0.9359 | 0.9541 | 0.9102 | 0.9454 | **0.9542** | 0.9152 |
| MR | 0.9208 | 0.9248 | 0.9211 | 0.9453 | **0.9513** | 0.9094 |
| ⟨AUC⟩ | 0.9029 | **0.9148** | 0.8513 | 0.8944 | 0.9102 | 0.8973 |

## IV. Discussions

In this paper, we propose two enhancement frameworks, CF and SCF, for local similarity indices. Extensive experiments indicate that CF-enhanced indices perform much better than the original ones, and SCF-enhanced indices can still significantly improve the performance of CF-enhanced indices. To our surprise, SCF-enhanced indices, like $\ddot{S}^{RA}$, can achieve competitive performance compared with state-of-the-art global algorithms, like SPM and LO. Notice that, although $\ddot{S}^{RA}$ itself can be treated as a novel local similarity index, what we propose is not one or a few novel indices but general frameworks to enhance local indices, so that we believe what proposed in this paper is of higher applicability.

We use the term *collaborative filtering* because we are inspired by user-based collaborative filtering techniques, however, the current methods are not same to collaborative filtering because in addition to the analogy, personalized recommendation and link prediction have some essential differences [2][8]. The proposed frameworks also contain similar ideas to linear oprimization [60] and transferring similarity [62], but the latter two eventually result in global indices that are more time-consuming.

**Acknowledgements**. This work was partially supported by the National Natural Science Foundation of China (Grant Nos. 11975071 and 61673086), the Science and Technology Department of Sichuan Province (Grant No. 2020YFS0007), the Chendu Science and Technology Agency (Grant No. 2020-YF05-00073-SN), and the Science Promotion Programme of UESTC, China (Grant No. Y03111023901014006).